# Large low-field magnetocaloric response in a ferromagnetic gadolinium orthophosphate

Ziyu W. Yang,[1, 2] Jie Zhang,[2, 3] Maocai Pi,[2, 3] Xubin Ye,[2] Chenxu Kang,[1] Xiaoliang Weng,[1] Wei Tang,[1] Hongzhi Cui,[1, *] Yu-Jia Zeng[1, *] and Youwen Long[2, 3, 4, *]

[1] *College of Civil and Transportation Engineering, College of Physics and Optoelectronic Engineering, Shenzhen University, Shenzhen, 518060 China. Email: h.z.cui@szu.edu.cn, yjzeng@szu.edu.cn.*

[2] *Beijing National Laboratory for Condensed Matter Physics, Institute of Physics, Chinese Academy of Sciences, Beijing, 100190 China. Email: ywlong@iphy.ac.cn.*

[3] *School of Physical Sciences, University of Chinese Academy of Sciences, Beijing, 100049 China.*

[4] *Songshan Lake Materials Laboratory, Dongguan, Guangdong, 523808 China.*

Bulk magnetic and thermodynamic measurements, along with mean-field calculations, were conducted on the ferromagnetic $K_3Gd_5(PO_4)_6$ powders. No magnetic ordering was observed until 2 K, while the application of an external field $B > 1$ T resulted in the splitting of the $Gd^{3+}$ ground state multiplet and induced a non-cooperative Schottky effect. The average nearest-neighbor exchange strength $|J_1/k_B|$ is determined to be 0.017 K, which leads to a remarkably large low-field magnetic entropy change $\Delta S_m$ = 36.2 J kg$^{-1}$ K$^{-1}$ under applied field change $B$ = 2 T at temperature $T$ = 2 K, as well as a maximum adiabatic temperature change $T_{ad}$ = 10.9 K. We contend that ferromagnetic gadolinium orthophosphates serve as a promising reservoir for exploring advanced magnetic refrigerants applicable under low magnetic fields.

## INTRODUCTION

Adiabatic demagnetization refrigeration (ADR), a compact solid cooling technique initially proposed by Langevin and Debye et al. at the beginning of the last century, offers an exceptionally appealing solution for cooling from several kelvin to the deep sub-kelvin regime.[1] It possesses remarkable attributes such as high reliability and efficiency, stable and precise temperature control, as well as minimal microphonics and thermophotonics effects.[2] This technique is gaining increasing significance in the fields of space detection, quantum computing, cold atom experiments, and gas liquefaction.[3]

The underlying principle of ADR is based on the magnetocaloric effect, which exploits the reduction in entropy resulting from the application of a magnetic field. In order to achieve enhanced cooling efficiency and reduce costs within commercially viable magnetic fields, typically below $B$ = 4 T, it is crucial to develop magnetocaloric materials with lower temperature operation limits and higher cooling power per unit mass or volume.[3b, 4] This necessitates specific requirements for magnetocaloric materials, including a low magnetic ordering transition temperature, a large magnetic moment, high thermal conductivity, and moderate robustness to crystal-field interactions. The compounds containing $Gd^{3+}$ and $Eu^{2+}$ with an $^8S_{7/2}$ ground state are thus highly appealing due to the potential for maximizing entropy utilization to $S_m = R\ln(2J+1)$, where $R$ is the gas constant, $J$ = 7/2, as exemplified by the benchmark refrigerant $Gd_3Ga_5O_{12}$, along with several other promising candidates such as $Gd(CHOO)_3$, $GdPO_4$, $GdF_3$, $Gd(OH)F_2$, $Ca_2GdSbO_6$, $Sr_2GdNbO_6$, $Gd_{9.33}Si_6O_{26}$, $EuTiO_3$ and $EuB_4O_7$, as well as molecular complexes [{$Gd(OAc)_3$-$(H_2O)_2$}$_2$]·4$H_2O$, $Gd_7$ cluster and {$Gd_{16}Na_2$} wheels.[4a, d, 5]

In addition to the aforementioned arguments, it is crucial to consider an appropriate magnetic ordering mechanism for optimizing overall refrigeration efficiency. Commencing from a conventional mean-field expression for free magnetic moments, the available magnetic entropy is,

$$\frac{S_m}{R} = \ln\sinh\left(J+\frac{1}{2}\right)x - \ln\sinh\left(\frac{x}{2}\right) - xJB_J(x) \quad (1)$$

where $B_J(x)$ represents the Brillouin function. As proposed by Wood et al, the variable $x$ can be expressed as,

$$x = \frac{g\mu_B}{k_B(T/T_o)}\left[B/T_o \pm \frac{3k_B B_J(x)}{g\mu_B(J+1)}\right] \quad (2)$$

where $g$ = 2, $k_B$ denotes the Boltzmann constant, and $B/T_o$ represents the applied field normalized to ordering temperature $T_o$ in units of tesla per kelvin.[6] The ± symbol represents systems with ferromagnetic (FM, +) and antiferromagnetic (AFM, -) correlations, whereas in paramagnets (PM), the variable $x$ is reduced to $x = g\mu_B(B/T_o)/[k_B(T/T_o)]$. Fig. 1 illustrates typical results obtained from mean-field calculations of the magnetic entropy change $\Delta S_m = S_m(0) - S_m(B)$, as a function of reduced temperature $T/T_o$ at low applied fields.

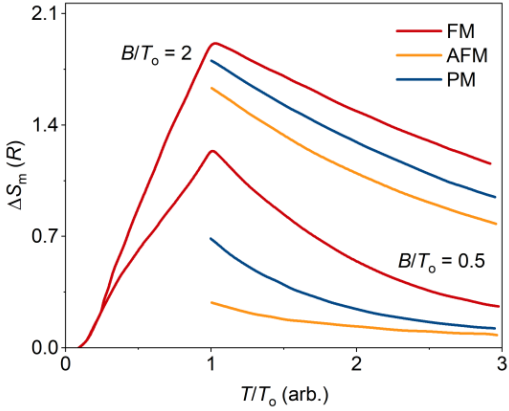

**Fig. 1.** Theoretical magnetic entropy changes $\Delta S_m$, normalized to the gas constant $R$, for $Gd^{3+}$ ions in configurations exhibiting paramagnetic, ferromagnetic and antiferromagnetic correlations. The temperature axis is dimensionless, represented as $T/T_o$, where $T_o$ represents the ordering temperature. The applied field $B/T_o$ is in units of tesla per kelvin.

The data below the Néel temperature for AFM systems are disregarded due to the difficulty in self-cooling an antiferromagnet.[7] Notably, significant differences can be observed among the three cases even at a temperature three times higher than $T_o$, indicating that ferromagnetic materials are more suitable for applications in low magnetic fields. This is supported by the remarkable $\Delta S_m$ values exhibited by the ferromagnetic $GdF_3$ (45.5 J kg$^{-1}$ K$^{-1}$ for $B$ = 2 T), and dimer complex [{Gd(OAc)$_3$-(H$_2$O)$_2$}$_2$]·4H$_2$O (27 J kg$^{-1}$ K$^{-1}$ for $B$ = 1 T).[5d, 8]

However, the stability of fluorides and water-containing molecular complexes remains uncertain, particularly in ultra-high-vacuum environments and applications requiring high-temperature baking, such as infrared space telescopes.[4c, 9] Therefore, it is highly desirable to explore other ferromagnetic materials that are stable without water. In this context, we present the previously unexplored magnetocaloric responses of the ferromagnetic gadolinium orthophosphate $K_3Gd_5(PO_4)_6$. Our comprehensive analysis of the compound's magnetic and thermodynamic properties demonstrates its significant potential as a candidate for magnetic cooling within the temperature range of 2 - 20 K, extending towards lower temperatures.

## MATERIALS AND METHODS

Polycrystalline $K_3Gd_5(PO_4)_6$, weighing approximately 3.0 g, was synthesized through solid-state reactions using stoichiometric amounts of commercially available starting materials: $K_2CO_3$ (>99.0%, TCI), $Gd_2O_3$ (99.999%, Adamas-beta, pre-heated at 1173 K), and $NH_4H_2PO_4$ (>99.0%, Energy Chemical). The reactants were thoroughly ground in an agate mortar and pressed into a $\varphi$-18 mm diameter pellet, which was then heated at 573 K for 6 h followed by heating at 873 K for 10 h in an alumina crucible with a lid. Subsequently, the pellet was pounded and re-compacted before being reacted at 1073 K for 30 h.

The powder X-ray diffraction patterns were collected using monochromated Cu $K\alpha$ radiation on a Huber diffraction and positioning equipment, and analyzed by Rietveld refinements with the GSAS-EXPGUI package.[10] The potassium mass fraction was determined using ICP-OES (Thermo Scientific iCAP PRO). The dc magnetization and susceptibility data were measured at applied field (0 to 7 T) and temperature (1.9 to 300 K) using a Quantum Design PPMS (PPMS-9). Caloric measurements were performed in the temperature range of 1.9 to 40 K, with varied applied fields of 0, 0.5, 1, 2, 5 and 7 T using the same experimental setup. To ensure optimal thermal contact, Apiezon N-grease was employed.

## RESULTS AND DISCUSSION

The powder X-ray diffraction patterns obtained at ambient temperature indicate the formation of pure phase products $K_3Gd_5(PO_4)_6$, which belong to the monoclinic system with space group $C2/c$. Rietveld refinements were conducted using an initial structural model reported by Zhu et al, resulting in residual factors of $wR_p/R_p$ = 2.13/1.57 = 1.36 and $\chi^2$ = 4.74, as well as refined unit cell parameters of $a$ = 17.4551(12) Å, $b$ = 6.9273(5) Å, $c$ = 18.0964 (12) Å and $\beta$ =114.4115(8) °.[11] The site occupancy was maintained constant, as confirmed by ICP-AES measurement revealing that potassium exhibited a mass fraction of 8.06 wt%, deviating only by 1% from the stoichiometric formula.

The sketched view in Fig. 2b illustrates the structural units of three crystallographically unique Gd (Gd1, Gd2, and Gd3) and two K (K1 and K2) atoms. Gd1 and Gd3 are each surrounded by four neighbouring Gd atoms, while Gd2 is surrounded by three neighbouring Gd atoms. The shortest distance between two adjacent Gd atoms (Gd1-Gd3) is approximately 3.9644(53) Å, whereas the longest distance measures 4.1444(44) Å, forming a three-dimensional framework of $Gd^{3+}$ cations with infinite tunnels along the $c$-axis where the K atoms are packed (represented as dark grey circles). The neighboring Gd-Gd distances fall within the typical range observed in other gadolinium orthophosphates, such as $GdPO_4$ ($P2_1/n$), which exhibits noncollinear antiferromagnetic ordering with a transition temperature $T_N = 0.77$ K.

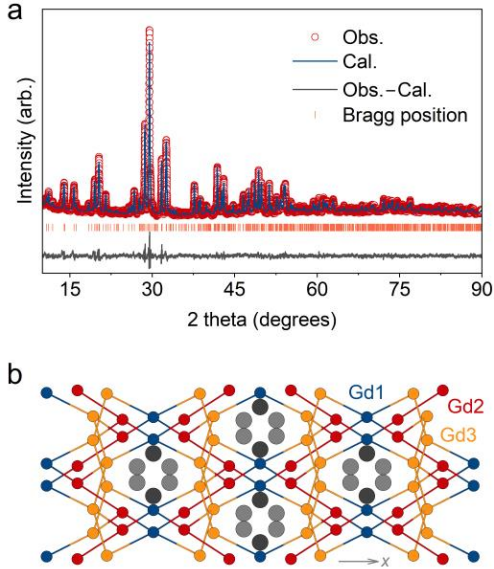

cooling to $T = 37$ K, below which it increases significantly and attains a magnitude of 10.6 $cm^3$ $mol^{-1}$ K at $T = 2$ K, indicative of the overall prevalence of ferromagnetic interactions (Fig. 3b). The $GdPO_4$, on the other hand, exhibits similar neighboring Gd-Gd distances while displaying antiferromagnetic ordering characterized by a propagation vector **k** = (1/2, 0, 1/2).[12]

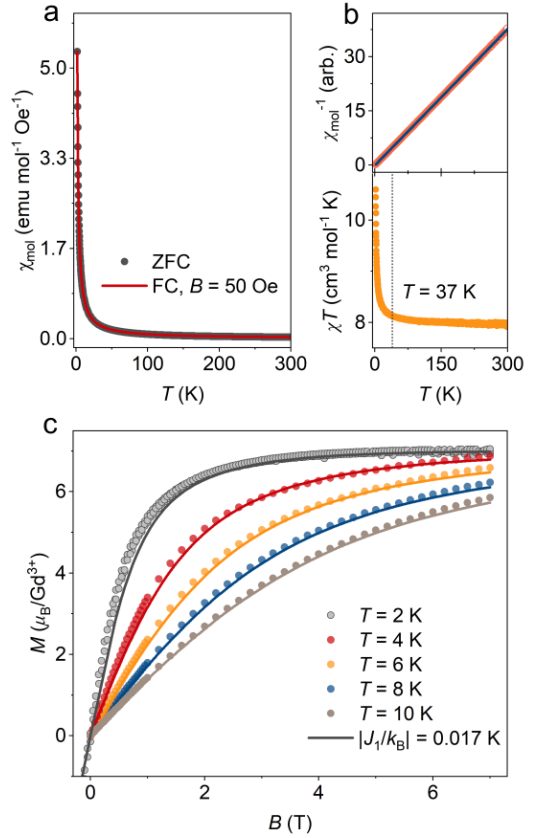

**Fig. 2.** (a) Room temperature powder X-ray diffraction pattern for $K_3Gd_5(PO_4)_6$, accompanied by the Rietveld refinement analysis. (b) Schematic representation of the crystal structure along the $c$-axis; K positions are denoted by dark grey circles.

**Fig. 3.** (a) Molar dc susceptibility $\chi_{mol}$ as a function of temperature, measured in an applied field $B = 50$ Oe, and presented as $\chi_{mol}T$ and $\chi_{mol}^{-1}$ (the solid line represents the Curie-Weiss fits) (b). (c) Isothermal magnetization $M$ as a function of applied field $B$ at various temperatures, accompanied by mean-field calculations using $|J_1/k_B|=0.017$ K (solid lines).

The dc magnetic susceptibility measured under a 50 Oe applied field is depicted in Fig. 3a, showing no indication of any ordering transition between $T = 2$ to 300 K. The zero-field-cooled (ZFC) data were fitted using a standard Curie-Weiss function with temperature-independent contribution, $\chi - \chi_0 = C/(T - \theta_W)$, resulting in a Weiss temperature of $\theta_W = 0.6$ K and an effective paramagnetic moment $\mu_{eff} = 7.99$ $\mu_B$, which are in broadly agreement with the free-ion moment of $g\sqrt{J(J+1)}J = 7.94$ $\mu_B$.

The molar $\chi T$ product exhibits a value of approximately 8.0 $cm^3$ $mol^{-1}$ K, consistent with that expected for free $Gd^{3+}$ ions, and displays only gradual enhancement upon

The isothermal magnetization $M(T, B)$ measured at $T = 2$ K and $B = 7$ T reaches its maximum value, $gJ\mu_B = 7\mu_B$, without exhibiting any noticeable magnetic hysteresis, validating the accessibility of the complete and reversable magnetic entropy (Fig. 3c). We then performed numerical simulations of the molar magnetization $M(T, B)$ curves using a mean-field approach with an average exchange coupling strength of $|J_1/k_B| = 0.017$ K, which yielded satisfactory agreement with experimental

values.[5c, 13] Given that $Gd^{3+}$ possesses an isotropic spin-only total angular momentum with $^8S_{7/2}$ ground state, the Curie-Weiss temperature can be realistically interpreted as $\theta_W = zJ_1S(S+1)/3 = 0.4$ K, where $z = 4$ denotes the nearest neighboring number, accurately capture the experimental susceptibility observables.

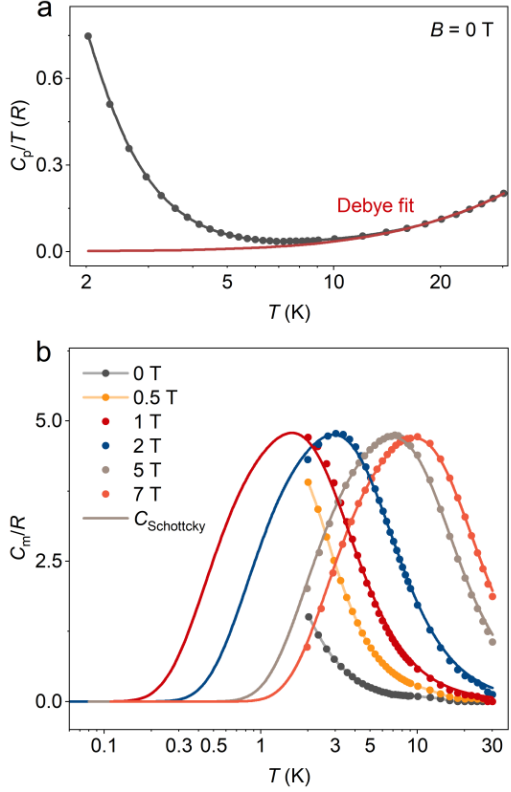

**Fig. 4.** (a) Molar heat capacity $C_p$, in nil field, normalized to the gas constant $R$. Solid line represents the lattice contribution utilizing a Debye polynomial. (b) Magnetic specific heat $C_m$, under applied fields of 0, 0.5, 1, 2, 5 and 7 T, and the model calculations using a two-sublattice Schottky anomaly (solid lines). The solid lines in the plots for $B = 0$ and 0.5 T are provided as visual aids only without Schottky fitting.

Fig. 4a depicts the molar heat capacity $C_p$ collected in nil field, where the field independent lattice phonon modes dominate above 8 K. This non-magnetic contribution is assumed to follow a simplified $T^3$ dependence, with additional $T^5$ and $T^7$ terms accounting for deviations from Debye behavior. The zero-field magnetic heat capacity $C_m$ (Fig. 4b) exhibits no evidence of magnetic ordering at $T = 2$ K, displaying a maximum value at 1.5 $R$/f.u., which is significantly smaller than the mean-field estimation of 2.4 $R/Gd^{3+}$. Upon increasing the applied magnetic field to $B > 1$ T, a broad feature emerges and the peak maximum shifts towards higher temperatures, indicating the presence of a Schottky anomaly associated with Zeeman splitting.[5b, 14] Starting from a general expression with $n$ discrete energy levels, $E_i$, as

$$C_{Sch}/R = \left\{\sum_{i=1}^{n} E_i^2 \exp(-E_i/T) \Big/ \sum_{i=1}^{n} \exp(-E_i/T) - \left[\sum_{i=1}^{n} E_i \exp(-E_i/T) \Big/ \sum_{i=1}^{n} \exp(-E_i/T)\right]^2\right\}/T^2 \quad (3)$$

the temperature-driven repopulation of the ground state sublevels of $Gd^{3+}$ can be produced very well, as shown in Fig. 4b.

Figure 5a illustrates the total entropy diagram $S_{tot} = S_L + S_m$, where $S_L$ represents the lattice entropy and $S_m$ corresponds to the magnetic entropy determined by integrating $C_m/T$ up to $T_{max} = 30$ K. To compensate for experimental limitations below $T < 2$ K, we extrapolated the heat capacity data using Schottky calculations for $B = 2, 5,$ and 7 T. For $B = 0, 0.5,$ and 1 T, a crude mean-field model was employed in the data consolidation process assuming that complete magnetic entropy is released at $T = 30$ K; this assumption is justified as $C_m$ approaches zero at $B = 0, 0.5,$ and 1 T (Fig. 4b).[2] The numerical calculation is performed as follows,

$$S_m(T,B) = S_m(T_{max},B) - \int_T^{T_{max}} \frac{C_m(\tau,B)}{\tau} d\tau, T \in [T_{min}, T_{max}] \quad (4)$$

where $S_m(T_{max}, B)$ represents the theoretical magnetic entropy with $J_1/k_B = 0.017$ K, and $T_{min} = 2$ K corresponding to the minimum temperature in the heat capacity measurement.

The so-obtained nil field entropy becomes temperature independent and approaches $S_L+5R\ln 8$ above $T = 6$ K, in agreement with the fully degenerate ground state multiplet of $S = 7/2$, as shown in Fig. 5a. The lattice entropy $S_L$ at $T = 20$ K is sufficiently small compared to the magnetic entropy $S_m$ at the same temperature, with a ratio of $S_L/S_m = 0.08$, which is lower than that of $GdPO_4$ ($S_L/S_m = 0.42$).[2] This enables more simplified Carnot cycles to operate below 20 K, as compared to the Stirling cycles that necessitate regenerations with a high $S_L/S_m$ ratio.

We now turn to the indirect estimation of the magnetic entropy change $\Delta S_m$, which can be readily derived from the experimental entropy diagram, as depicted schematically by the arrows in Fig. 5a. Here, the shaded area represents an ideal Carnot cycle, while $S_A - S_B$ denotes the magnetic entropy change resulting from a $\Delta B = 5$ T applied field variation at $T = T_{hot}$. The evaluation can also be performed utilizing Maxwell's relation based on the isothermal magnetization curves, as

$$\Delta S_m(T,B) = \int_0^B \frac{\partial M(T,B')}{\partial T} dB' \quad (5)$$

with the corresponding outcomes depicted as open circles in Fig. 5b. These results are consistent with each other, thereby further validating that our experimental uncertainties do not compromise the estimation of the magnetocaloric response.

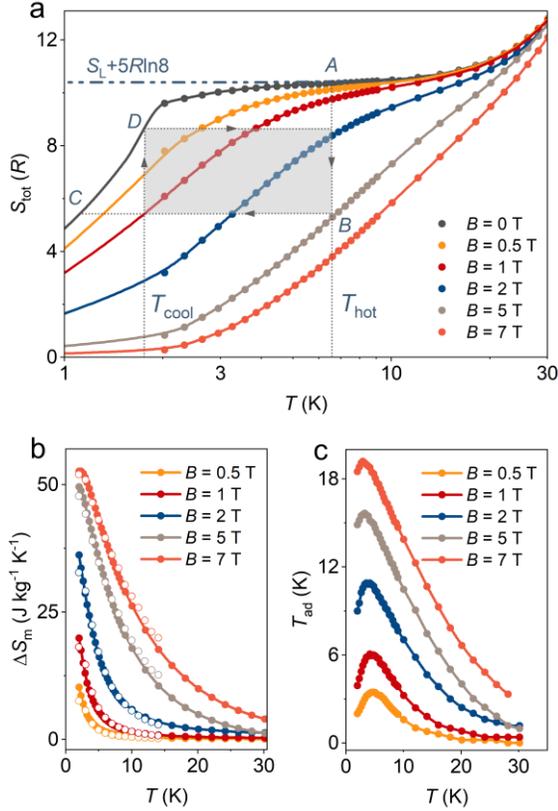

**Fig. 5.** (a) Temperature depended molar entropy $S_{tot}$, for several applied fields determined by integrating the heat capacity data. The shaded region represents an ideal Carnot cycle starting from an initial condition $T_{hot}$ = 6.6 K and $B$ = 5 T, ending at a final temperature $T_{cool}$ = 1.8 K. The arrows represent the magnetic entropy change, $\Delta S_m$ = $S_A \rightarrow S_B$, corresponding to an applied field variation of 0 $\rightarrow$ 5 T. Calculated results are depicted in (b), while the open circles are estimated using Maxwell relations derived from experimental magnetization data. The adiabatic demagnetization cooling process, denoted as $B$ to $C$ in (a), represents the ideal scenario that yields the theoretical adiabatic temperature change $T_{ad}$, as depicted in (c).

The magnetic entropy change $\Delta S_m$ is 52.6 J kg$^{-1}$ K$^{-1}$ (0.26 J cm$^{-3}$ K$^{-1}$) at $T$ = 2 K for an applied field change of 7 T, achieving nearly 90% of the maximum available entropy value (5$R$ln8). Additionally, it exhibits values of 36.2 (0.18 J cm$^{-3}$ K$^{-1}$) and 19.9 J kg$^{-1}$ K$^{-1}$ (0.10 J cm$^{-3}$ K$^{-1}$) under applied field changes $\Delta B$ = 2 and 1 T at the same temperature, respectively, surpassing the performance of the benchmark refrigerant agent gadolinium gallium garnet Gd$_3$Ga$_5$O$_{12}$ (GGG) under $\Delta B$ = 2 T ($\Delta S_m$ = 20.5 J kg$^{-1}$ K$^{-1}$ or 0.15 J cm$^{-3}$ K$^{-1}$).[4d, 5k] These values exhibit significant importance in comparison to the majority of magnetocaloric materials documented in the literature as well (Table. 1), thereby confirming the potential viability of K$_3$Gd$_5$(PO$_4$)$_6$ for implementation in adiabatic demagnetization refrigerators operating within the temperature range of 2-20 K and extending towards lower temperatures.

**Table 1.** Magnetic entropy changes of some top-performing magnetocaloric materials under a 2 T applied field variation at $T$ = 2 K.

| Materials | -$\Delta S_m$ [J kg$^{-1}$ K$^{-1}$] | $\Delta B$ [T] | Ref. |
|---|---|---|---|
| Gd$_{9.33}$Si$_6$O$_{26}$ | 27.8 | 2 | [5b] |
| Gd$_2$SiO$_5$ | 11.4 | 2 | [4e] |
| Gd(HCOO)$_3$ | 34.8 | 2 | [4a] |
| {Gd$_{16}$Na$_2$} wheel | 36.1 | 2 | [5e] |
| GdVO$_4$ | 10.5 | 3 | [15] |
| EuB$_4$O$_7$ | 31.3 | 2 | [5f] |
| Gd$_3$CrGa$_4$O$_{12}$ | 16.6 | 2 | [4d] |
| Gd(OH)F$_2$ | 45.2 | 2 | [5g] |
| Gd$_{152}$Ni$_{114}$@Cl$_{24}$ cluster | 29.2 | 3 | [16] |
| Gd$_3$BWO$_9$ | 24.4 | 2 | [17] |
| GdLiF$_4$ | 35.2 | 2 | [18] |
| K$_3$Gd$_5$(PO$_4$)$_6$ | 36.2 | 2 | This work |

The rough calculation of the theoretical adiabatic temperature change $T_{ad}$ during a cooling process is depicted in Fig. 5a, where $T_{ad}$ = $T_B$ ($T_{hot}$)-$T_C$. By applying a field change of 7 T, the maximum achievable $T_{ad}$ is found to be 19.2 K, while field changes of 2 and 1 T result in temperature changes of 10.9 and 6.1 K respectively (Fig. 5c). It should be noted that in the Carnot process, the final state is typically set at a non-zero field value, implying that demagnetization is expected to cool the material down to $T_{cool}$ ($T_{hot}$ - $T_{cool}$ < $T_{ad}$), as illustrated in Fig. 5a. Besides, both indirect estimation methods rely on numer-ical integrations and data extrapolations, which inherently introduce the possibility of calculation errors. Therefore, it is advisable to employ direct measuring approaches for assessing the actual cooling performances.[3c, 4c, 5i, 19]

## CONCLUSIONS

In summary, the magnetocaloric effect is evaluated in the ferromagnetic $K_3Gd_5(PO_4)_6$, which exhibits moderate magnetic correlations among the $Gd^{3+}$ centers. The title compound exhibits no evidence of magnetic ordering until $T = 2$ K, and a non-cooperative Schottky effect emerges when a magnetic field $B > 1$ T is applied.

A simplified mean-field approach was employed to facilitate the analysis of the experimental magnetization and entropy diagram. The material demonstrates a significant magnetocaloric response at low fields, with values of 36.2 J kg$^{-1}$ K$^{-1}$ (0.18 J cm$^{-3}$ K$^{-1}$) and 19.9 J kg$^{-1}$ K$^{-1}$ (0.10 J cm$^{-3}$ K$^{-1}$) under applied field changes $\Delta B = 2$ and 1 T at $T = 2$ K, respectively, surpassing the performance of the benchmark refrigerant GGG. The lattice entropy $S_L$ at $T = 20$ K is significantly smaller compared to the magnetic entropy $S_m$, making it an attractive candidate for utilization in both Carnot and Stirling cycles within the temperature range of 2 – 20 K, as well as at even lower temperatures. The research conducted here indicates an interest in exploring and manipulating other ferromagnetic phosphates for the purpose of magnetic cooling, including those that are magnetically doped to generate internal polarizing fields.

## AUTHOR CONTRIBUTIONS

Z.Y.: conceptualization, investigation, formal analysis, writing─original draft, and funding acquisition. J.Z., M.P., X.Y., C.K., X.W. and W.T.: investigation. H.C., Y.J.Z., and Y.L.: supervision and funding acquisition.

## ACKNOWLEDGEMENT

The research here was supported by the National Key R&D Program of China (2021YFA1400300), the Beijing Natural Science Foundation (Z200007), the Guangdong Basic and Applied Basic Research Foundation (2022A1515111009), the National Natural Science Foundation of China (51925804, 52273298 and 11934017), the Shenzhen Science and Technology Program (JCYJ20210324095611032), and the Chinese Academy of Sciences (XDB33000000).

## REFERENCES

[1] a) P. J. Shirron, *Cryogenics*, **2014**, 62, 130; b) P. Wikus, E. Canavan, S. T. Heine, K. Matsumoto, T. Numazawa, *Cryogenics*, **2014**, 62, 150; c) A. Alahmer, M. Al-Amayreh, A. O. Mostafa, M. Al-Dabbas, H. Rezk, *Energies*, **2021**, 14, 4662; d) N. A. de Oliveira, P. J. von Ranke, *Phys. Rep*, **2010**, 489, 89.

[2] J. A. Barclay, W. A. Steyert, *Cryogenics*, **1982**, 22, 73.

[3] a) P. J. S. Amir E. Jahromi, Michael J. DiPirro, in *Cryogenic Engineering Conference and International Cryogenic Materials Conference (CEC/ICMC)*, 2019; b) D. A. P. Brasiliano, J.-M. Duval, C. Marin, E. Bichaud, J.-P. Brison, M. Zhitomirsky, N. Luchier, *Cryogenics*, **2020**, 105, 103002; c) B. Wolf, Y. Tsui, D. Jaiswal-Nagar, U. Tutsch, A. Honecker, K. Remović-Langer, G. Hofmann, A. Prokofiev, W. Assmus, G. Donath, *Proc. Natl. Acad. Sci.*, **2011**, 108, 6862.

[4] a) G. Lorusso, J. W. Sharples, E. Palacios, O. Roubeau, E. K. Brechin, R. Sessoli, A. Rossin, F. Tuna, E. J. McInnes, D. Collison, M. Evangelisti, *Adv Mater*, **2013**, 25, 4653; b) Z. Mo, J. Gong, H. Xie, L. Zhang, Q. Fu, X. Gao, Z. Li, J. Shen, *Chinese Phys. B*, **2023**, 32, 027503; c) Y. Tokiwa, S. Bachus, K. Kavita, A. Jesche, A. A. Tsirlin, P. Gegenwart, *Commun. Mater.*, **2021**, 2, 42; d) N. K. Chogondahalli Muniraju, R. Baral, Y. Tian, R. Li, N. Poudel, K. Gofryk, N. Barisic, B. Kiefer, J. H. Ross, Jr., H. S. Nair, *Inorg. Chem.*, **2020**, 59, 15144; e) Z. W. Yang, J. Zhang, D. Lu, X. Zhang, H. Zhao, H. Cui, Y.-J. Zeng, Y. Long, *Inorg. Chem.*, **2023**, 62, 5282.

[5] a) M. Evangelisti, O. Roubeau, E. Palacios, A. Camón, T. N. Hooper, E. K. Brechin, J. J. Alonso, Wiley-VCH, 2011; b) Z. W. Yang, J. Zhang, B. Liu, X. Zhang, D. Lu, H. Zhao, M. Pi, H. Cui, Y. J. Zeng, Z. Pan, *Adv. Sci*, **2024**, 2306842; c) E. C. Koskelo, C. Liu, P. Mukherjee, N. D. Kelly, S. E. Dutton, *Chem. Mater.*, **2022**, 34, 3440; d) Y.-C. Chen, J. Prokleška, W.-J. Xu, J.-L. Liu, J. Liu, W.-X. Zhang, J.-H. Jia, V. Sechovský, M.-L. Tong, *J. Mater. Chem. C*, **2015**, 3, 12206; e) T. G. Tziotzi, D. Gracia, S. J. Dalgarno, J. Schnack, M. Evangelisti, E. K. Brechin, C. J. Milios, *J. Am. Chem. Soc.*, **2023**, 145, 7743; f) Y. Wang, J. Xiang, L. Zhang, J. Gong, W. Li, Z. Mo, J. Shen, *J. Am. Chem. Soc.*, **2024**; g) Q. Xu, B. Liu, M. Ye, G. Zhuang, L. Long, L. Zheng, *J. Am. Chem. Soc.*, **2022**, 144, 13787; h) P. Xu, Z. Ma, P. Wang, H. Wang, L. Li, *Mater. Today Phys.*, **2021**, 20, 100470; i) J. W. Sharples, D. Collison, E. J. L. McInnes, J. Schnack, E. Palacios, M. Evangelisti, *Nat. Commun.*, **2014**, 5, 5321; j) A. Midya, P. Mandal, K. Rubi, R. Chen,

J.-S. Wang, R. Mahendiran, G. Lorusso, M. Evangelisti, *Phys. Rev. B*, **2016**, 93, 094422; k) P. Mukherjee, A. C. Sackville Hamilton, H. F. J. Glass, S. E. Dutton, *J. Phys.: Condens. Matter*, **2017**, 29, 405808.

[6] M. Wood, W. Potter, *Cryogenics*, **1985**, 25, 667.

[7] P. Wikus, G. Burghart, E. Figueroa-Feliciano, *Cryogenics*, **2011**, 51, 555.

[8] M. Evangelisti, O. Roubeau, E. Palacios, A. Camón, T. N. Hooper, E. K. Brechin, J. J. Alonso, *Angew. Chem. Int. Ed.*, **2011**, 50, 6606.

[9] M. DiPirro, E. Canavan, P. Shirron, J. Tuttle, *Cryogenics*, **2004**, 44, 559.

[10] B. Toby, *J. Appl. Crystallogr.*, **2001**, 34, 210.

[11] a) J. Zhu, W.-D. Cheng, D.-S. Wu, H. Zhang, Y.-J. Gong, H.-N. Tong, D. Zhao, *Cryst. Growth Des.*, **2006**, 6, 1649; b) S. Bevara, S. N. Achary, K. K. Mishra, T. Ravindran, A. K. Sinha, P. Sastry, A. K. Tyagi, *Phys. Chem. Chem. Phys.*, **2017**, 19, 6030.

[12] E. Palacios, J. A. Rodríguez-Velamazán, M. Evangelisti, G. J. McIntyre, G. Lorusso, D. Visser, L. J. De Jongh, L. A. Boatner, *Phys. Rev. B*, **2014**, 90, 214423.

[13] a) Z. W. Yang, S. Qin, J. Zhang, D. Lu, H. Zhao, C. Kang, H. Cui, Y. Long, Y.-J. Zeng, *Mater. Today Phys.*, **2022**, 27, 100810; b) C. Wellm, J. Zeisner, A. Alfonsov, M. I. Sturza, G. Bastien, S. Gaß, S. Wurmehl, A. U. B. Wolter, B. Büchner, V. Kataev, *Phys. Rev. B*, **2020**, 102, 214414

[14] A. Jesche, N. Winterhalter-Stocker, F. Hirschberger, A. Bellon, S. Bachus, Y. Tokiwa, A. A. Tsirlin, P. Gegenwart, *Phys. Rev. B*, **2023**, 107, 104402.

[15] E. Palacios, M. Evangelisti, R. Sáez-Puche, A. J. Dos Santos-García, F. Fernández-Martínez, C. Cascales, M. Castro, R. Burriel, O. Fabelo, J. A. Rodríguez-Velamazán, *Phys. Rev. B*, **2018**, 97, 214401.

[16] N. Xu, W. Chen, Y.-S. Ding, Z. Zheng, *J. Am. Chem. Soc.*, **2024**, 146, 9506.

[17] Z. Yang, H. Zhang, M. Bai, W. Li, S. Huang, S. Ruan, Y.-J. Zeng, *J. Mater. Chem. C*, **2020**, 8, 11866.

[18] T. Numazawa, K. Kamiya, P. Shirron, M. DiPirro, K. Matsumoto, *AIP Conference Proceedings*, **2006**, 850, 1579.

[19] D. Jang, T. Gruner, A. Steppke, K. Mitsumoto, C. Geibel, M. Brando, *Nat. Commun.*, **2015**, 6, 8680.